\documentclass[proof]{pasj00}     
\SetRunningHead{Y. Takagi, Y. Itoh, and Y. Oasa}{Age determination method of PMS stars with spectroscopy}
\Received{}
\Accepted{}
\Published{}         
\begin{document}
\title{Age Determination Method of Pre-Main Sequence Stars with High-Resolution {\it I}-Band Spectroscopy\thanks{Based in part on data collected at Subaru Telescope, which is operated by the National Astronomical Observatory of Japan.}\thanks{This research has made use of the SIMBAD database, operated at CDS, Strasbourg, France.}} 
\author{Yuhei \textsc{Takagi}, Yoichi \textsc{Itoh}}
\affil{Graduate School of Science, Kobe University, 1-1 Rokkodai, Nada, Kobe, Hyogo 657-8501}
\email{takagi@stu.kobe-u.ac.jp}
\and
\author{Yumiko \textsc{Oasa}}
\affil{Faculty of Education, Saitama University, 255 Shimo-Okubo, Sakura, Saitama, Saitama 338-8570}
\maketitle

\begin{abstract}
We present a new method for determining the age of late-K type pre-main sequence (PMS) stars by deriving their surface gravity from high-resolution {\it I}-band spectroscopy. Since PMS stars contract as they evolve, age can be estimated from surface gravity. We used the equivalent width ratio (EWR) of nearby absorption lines to create a surface gravity diagnostic of PMS stars that is free of uncertainties due to veiling. The ratios of Fe (8186.7$\AA$ and 8204.9$\AA$) and Na (8183.3$\AA$ and 8194.8$\AA$) absorption lines were calculated for giants, main-sequence stars, and weak-line T Tauri stars. Effective temperatures were nearly equal across the sample. The Fe to Na EWR (Fe/Na) decreases significantly with increasing surface gravity, denoting that Fe/Na is a desirable diagnostic for deriving the surface gravity of pre-main sequence stars. The surface gravity of PMS stars with 0.8 $M_{\odot}$ is able to be determined with an accuracy of 0.1-0.2, which conducts the age of PMS stars within a factor of 1.5, in average.
\end{abstract}

\section{Introduction}
The age of pre-main sequence (PMS) stars is a vital parameter for studying the evolution of young stars. Accurate ages are needed to describe the evolutions of both the photosphere (e.g., rotational processes, chemical evolution) and the circumstellar material (e.g., circumstellar disk, protoplanets, outflows). Determining the age of PMS stars lets us study the evolution of the stars themselves and the history of star formation in their star-forming region (e.g., \cite{Palla2002}). Age is usually determined by comparing the luminosity and the effective temperature ($T_{\mathrm{eff}}$) estimated from broadband photometry or spectroscopy with stellar evolution tracks on the Hertzsprung-Russell (H-R) diagram (\cite{Cohen1979}; \cite{Strom1989}; \cite{Kenyon1995}). However, a precise estimate of PMS star luminosity is difficult. This difficulty arises from extinction, uncertainties in the distance measurements, and continuum excess caused from accretion and the heated circumstellar disk (e.g., veiling). These uncertainties may lead to an incorrect luminosity value. 

The age of a PMS star is able to be determined by estimating the surface gravity ($g$) instead of the luminosity. The surface gravity of a PMS star increases with time because of photospheric contraction. Spectroscopic observations are suitable for determining stellar parameters such as $T_{\mathrm{eff}}$, log $g$, metal abundance, and radial velocity, since the absorption line profile depends on these parameters. Although the spectra of PMS stars are veiled by the excess, the equivalent width ratio (EWR) of absorption lines is less contaminated by veiling \citep{Meyer1998}. \citet{Meyer1998} identified the EWR of OH (1.6892$\mu$m) and Mg (1.5760$\mu$m) as a diagnostic of $T_{\mathrm{eff}}$. They also used the EWR of CO (1.6207$\mu$m + 1.6617$\mu$m) and Mg (1.5760$\mu$m) as a luminosity indicator (which reflects the surface gravity). \citet{Doppmann2003} derived the $T_{\mathrm{eff}}$, radial velocity, and the amount of veiling by fitting observational 2.2$\mu$m spectra with synthesized spectra. They estimated log $g$ by calculating the EWR of the Na interval at 2.2$\mu$m (including the absorption lines of Na, Sc, Si) and the CO bandhead at 2.3$\mu$m. When the EWR is calculated by taking the ratio of absorption lines separated, a correction for the amount of veiling is necessary, as it is wavelength dependent. 

To derive the EWR directly from observed spectra without any veiling correction, we calculate EWRs using closely spaced absorption lines. This is practicable because the effects of veiling are nearly constant over small wavelength ranges. We present a new method for determining the age of mid- to late-K type PMS stars using EWRs of nearby absorption lines obtained from high-resolution {\it I}-band spectroscopy. We observed 30 giant stars, 6 main-sequence stars, and 4 weak-line T Tauri stars (WTTSs) with $T_{\mathrm{eff}}$ of around 4200K to derive a relationship between log $g$ and the EWRs of Fe (8186.7$\AA$, 8204.9$\AA$) and Na (8183.3$\AA$, 8194.8$\AA$).

\section{Method}
To make highly accurate determinations of PMS star age, precise log $g$ estimation is necessary. We selected the wavelength range in {\it I}-band (8180$\AA$ to 8210$\AA$), where strong atomic lines exist. This wavelength is suitable for low-mass PMS stars of which $T_{\mathrm{eff}}$ is around 4000K. Such PMS stars in the Hayashi phase \citep{Hayashi1961} evolve with roughly constant $T_{\mathrm{eff}}$. Therefore, their age may be approximately determined from log $g$. 

The absorption lines used in the EWR must be selected with care to minimize uncertainty in the resultant log $g$ value. We selected the $I$-band Na lines (8183.3$\AA$, 8194.8$\AA$) and Fe lines (8186.7$\AA$ (blended with vanadium), 8204.9$\AA$). These lines are sensitive to both $T_{\mathrm{eff}}$ and log $g$. Na lines are one of the strongest lines in the {\it I}-band. They are reached to the limiting depth, and the wing of these absorption lines is pressure-broadened in stars with large $g$. Fe lines are also strong lines in the $I$-band, but they are not reached to the limiting depth. The Fe line strength decreases with increasing $g$, depending on the growth of continuum opacity. Therefore, EWRs calculated from Na and Fe absorption line pair (Fe/Na) are considered suitable for deriving stellar surface gravities. In addition, assuming the abundance ratios of the elements are equal, EWR has no dependence on metal abundance. 

However, the relation between Fe/Na and surface gravity is yet unknown. To derive this relationship, a high-resolution spectroscopic observation of stars with fixed $T_{\mathrm{eff}}$ and varying log $g$ was necessary. Since the surface gravities of PMS stars are between that of giants and main-sequence stars, we observed giants, main-sequence stars. Also the WTTSs were observed to verify the effectiveness of this method to PMS stars.

\section{Observations and Data Reduction}

\subsection{Object Properties}

The selected sample objects and their properties are listed in table 1. We selected 30 giants (including subgiants) and 6 main-sequence stars to estimate the relationship between $g$ and the Fe/Na. All targets are single stars. 25 giants and 4 main-sequence stars have $T_{\mathrm{eff}}$ of 4200-4300K, since the radiation peak of these stars is in the {\it I}-band and they are bright enough to observe. Five giants and two dwarfs with $T_{\mathrm{eff}}$ of 4600-4700K were included to provide an estimate of the dependence of EWRs on $T_{\mathrm{eff}}$. 

To derive the surface gravity in advance, giants and main-sequence stars of which parallax was measured by the Hipparcos satellite \citep{Perryman1997} were selected. The log $g$ of these objects was calculated as follows:
\begin{center}
\begin{equation}
\log g = \log \frac{M}{M_{\odot}} + 4\log \frac{T}{T_{\odot}} - \log \frac{L}{L_{\odot}} + \log g_{\odot},
\end{equation}
\end{center}
where $M$ is the mass, $T$ the effective temperature, and $L$ the luminosity. The luminosity was calculated from the $V$-band magnitude given in SIMBAD (taken from previous photometric studies), and the parallax. For the effective temperature, the $T_{\mathrm{eff}}$-spectral type relationship \citep{Lang1992} was used. The error of the effective temperature is not included in the log $g$ estimation and therefore the error of the log $g$ may be underestimated. The relationship of $M$ and spectral type in \citet{Drilling2000} was used to derive the masses of the main-sequence stars. The giant star masses were estimated by comparing luminosities and temperatures with the evolutionary tracks given in \citet{Lejeune2001} (figure 1). Since giants smaller than 2.0 $M_{\odot}$ evolve rapidly after the helium flash, we used the pre-flash evolutionary track for these stars. The derived log $g$ of $\beta$ Gem (2.73$\pm$0.01) is consistent with the result of a spectroscopic study \citep{Drake1991}, in which it was estimated as 2.75$\pm$0.15. The large error in the surface gravity of the giants is attributable to the errors in luminosity (figure 1), which arise from the parallax errors.

Four WTTSs belonging to the Taurus-Auriga star-forming region were also added to our samples to test the effectiveness of the derived EWR-log $g$ relationship. The log $g$ of these WTTSs were also estimated from equation (1). Temperatures were estimated using \citet{Lang1992}, following the procedure for the main-sequence stars. We derived luminosities from photometry, including corrections for extinction ($A_V$), since the excess from the circumstellar disk is small. The $A_V$ for HBC 374 and V827 Tau were taken from \citet{Kenyon1995}. For the other objects, RX J0452.5+1730 and RX J0459.7+1430, we estimated the $A_V$ by fitting the SED corrected by an arbitrary $A_V$ to the $V$, $J$, $H$, and $K$-band magnitudes. The $V$-band magnitude was taken from SIMBAD, and the 2MASS catalog was used for the $J$, $H$, and $K$-band magnitudes. The estimated $A_V$ was 0.69 and 0.67 for RX J0452.5+1730 and RX J0459.7+1430, respectively. The mass of each star was estimated by using the H-R diagram, comparing the temperature and luminosity with the evolutionary model of \citet{Baraffe1998}. The mass of HBC 374 and V827 Tau were 0.8 $M_{\odot}$. On the other hand, the masses of RX J0452.5+1730 and RX J0459.7+1430 were 1.20 $M_{\odot}$ and 1.25 $M_{\odot}$, respectively. The distance to the Taurus-Auriga molecular cloud was fixed at 142$\pm$14pc \citep{Wichmann1998}. 

\begin{figure}
   \begin{center}
      \FigureFile(80mm,78.6mm){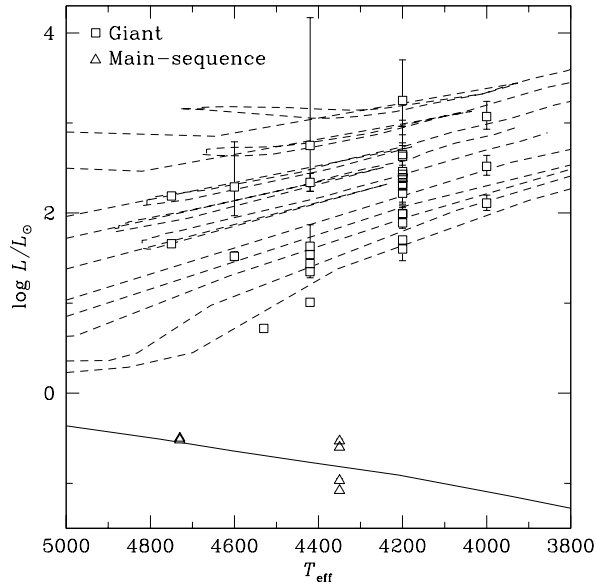}
      \end{center}
   \caption{
   The H-R diagram for the giants and main-sequence stars. The dashed lines are the evolutionary tracks from \citet{Lejeune2001}; $M$ = 0.90, 1.00, 1.25, 1.50, 1.70, 2.00, 2.50, 3.00, 4.00, and 5.00 $M_{\odot}$ from the bottom. The solid line indicates the isochrone for $10^9$yr \citep{Baraffe1998}.
   }\label{somelabel}
\end{figure}

\begin{longtable}{lcrrr}
   \caption{The list of the observed stars and their properties.}
\hline 
\multicolumn{1}{c}{Name} & Sp. Type & $T_{\mathrm{eff}}$ ($K$) & log $g$ ($g$: cm s$^{-2}$) & S/N \\
\hline
\endhead
\hline
\endfoot
\hline
\endlastfoot
Giants &  &  &  &  \\
\hline
   $\beta$ Gem	& K0IIIb	& 4750	& 2.73$\pm$0.01				&	160	\\
   HD 19656	& K0III		& 4750	& 2.38$\pm$0.02 			&	100	\\
   HD 76291	& K1IV		& 4600	& 2.72$^{+0.02}_{-0.04}$		&	170	\\
   HD 100204	& K1IV		& 4600	& 2.01$^{+0.19}_{-0.25}$		&	140	\\
   HD 145148	& K1.5IV	& 4530	& 3.23$\pm$0.03 			&	170	\\
   HD 45512	& K2III-IV	& 4420	& 2.55$\pm$0.03 			&	130	\\
   HD 101978	& K2IV		& 4420	& 1.82$^{+0.27}_{-1.07}$		&	160	\\
   HD 106102	& K2III-IV	& 4420	& 2.61$^{+0.08}_{-0.05}$		&	140	\\
   HD 108299	& K2IV		& 4420	& 2.46$^{+0.11}_{-0.13}$		&	100	\\
   HD 110501	& K2III-IV	& 4420	& 2.91$\pm$0.04 			&	110	\\
   HD 121146	& K2IV		& 4420	& 2.50$\pm$0.02 			&	190	\\
   HD 147142	& K2IV		& 4420	& 2.03$^{+0.06}_{-0.03}$		&	110	\\
   HD 49738	& K3III		& 4200	& 1.33$^{+0.20}_{-0.25}$		&	110	\\
   HD 65759	& K3III		& 4200	& 1.66$^{+0.09}_{-0.11}$		&	110	\\
   HD 86369	& K3III		& 4200	& 1.89$^{+0.06}_{-0.09}$		&	160	\\
   HD 88231	& K3III		& 4200	& 1.76$^{+0.10}_{-0.06}$		&	80	\\
   HD 102328	& K3III		& 4200	& 2.15$\pm$0.02 			&	90	\\
   HD 109012	& K3III-IV	& 4200	& 2.28$^{+0.12}_{-0.14}$		&	160	\\
   HD 113637	& K3III		& 4200	& 1.66$^{+0.11}_{-0.18}$		&	120	\\
   HD 115478	& K3III		& 4200	& 2.01$\pm$0.03 			&	140	\\
   HD 118839	& K3III		& 4200	& 1.83$\pm$0.10 			&	130	\\
   HD 129245	& K3III		& 4200	& 2.02$^{+0.03}_{-0.02}$		&	90	\\
   HD 132304	& K3III		& 4200	& 1.73$^{+0.16}_{-0.13}$		&	130	\\
   HD 137688	& K3III		& 4200	& 1.79$^{+0.17}_{-0.18}$		&	120	\\
   HD 146388	& K3III		& 4200	& 2.06$^{+0.01}_{-0.04}$		&	140	\\
   HD 156093	& K3III		& 4200	& 1.75$^{+0.10}_{-0.06}$		&	170	\\
   HD 163547	& K3III		& 4200	& 1.83$^{+0.05}_{-0.06}$		&	190	\\
   HD 10824	& K4III		& 4000	& 1.52$^{+0.08}_{-0.09}$		&	150	\\
   HD 20644	& K4III		& 4000	& 1.27$^{+0.04}_{-0.08}$		&	100	\\
   HD 23413	& K4III		& 4000	& 1.68$^{+0.08}_{-0.00}$		&	120	\\
   \hline
   Main-sequence stars & & & & \\
   \hline
   GL 105	& K3V		& 4730	& 4.46$\pm$0.01				&	270 	\\
   GL 183	& K3V		& 4730	& 4.44$\pm$0.01				&	110 	\\
   GL 141	& K5V		& 4350	& 4.37$\pm$0.01				&	120 	\\
   GL 380	& K5V		& 4350	& 4.85$\pm$0.00				&	220 	\\
   GL 397	& K5		& 4350	& 4.73$\pm$0.02				&	130 	\\
   GJ 3678	& K5V		& 4350	& 4.30$\pm$0.02				&	150 	\\
   \hline
   WTTS & & & & \\
   \hline
   RX J0452.5+1730 & K4		& 4590	& 4.22$^{+0.05}_{-0.07}$		&	60	\\
   RX J0459.7+1430 & K4		& 4590	& 4.13$^{+0.07}_{-0.08}$		&	60	\\
   HBC 374	   & K7		& 4060	& 3.92$^{+0.08}_{-0.09}$		&	70	\\ 
   V827 Tau	   & K7		& 4060	& 3.92$^{+0.08}_{-0.10}$		&	70	\\ \hline
   \end{longtable}

\subsection{Observations}

High-resolution {\it I}-band spectroscopy of 30 giants and 6 main-sequence stars was carried out on 2007 January 17 to 24 and 2008 March 24 to 27 using the High Dispersion Echelle Spectrograph (HIDES) on the Okayama Astrophysical Observatory 1.88 m telescope. The slit width was set to 220$\mu$m, giving a resolution of $\sim$60000. The angle of the red cross-disperser was selected to center the 69th order of the echelle spectrograph (8165$\sim$8300$\AA$) on the 2048$\times$4096 CCD. The integration time range was 90$\sim$7200 sec, yielding a S/N of $\sim$100. The maximum integration time was set to 1800 sec to reduce the risk of cosmic rays. Exposures of faint objects were divided into several frames. Flat frame were taken using the flat lamp equipped in the instrument, and were obtained at the beginning and end of each night's observations. Bias frames and Th-Ar lamp frames for wavelength calibration were taken with the same frequency. 

The spectra of the 4 WTTSs were obtained using the Subaru Telescope with High Dispersion Spectrograph (HDS) on 2007 September 18. The slit width was set to 0.6" (R$\sim$60000). The instrument was set to the StdNIRb mode, to obtain the Na and Fe absorption lines in the 73rd order. The integration time for each object was 900 sec. Flat, bias, and Th-Ar frames were also obtained. 

\subsection{Data Reduction}

The data were reduced using Image Reduction and Analysis Facility (IRAF) software package\footnote{IRAF is distributed by the National Optical Astronomy Observatory.}. The same procedures were applied to both HIDES data and HDS data. First, the bias was removed for each frame by subtracting the average level of the overscan area of each frame, and also by subtracting the averaged bias frame. Flat fielding was accomplished with the normalized flat frame. Cosmic rays and scattered light were removed. Next, the spectrum was extracted from each object frame. The comparison data obtained with the Th-Ar lamp were used for wavelength calibration. When the object was taken several times, the spectra were combined after wavelength calibration to improve the signal-to-noise ratio. The spectra were normalized at the continuum level in the last procedure. Spectra samples are displayed in figure 2.

\begin{figure}
   \begin{center}
      \FigureFile(80mm,71.9mm){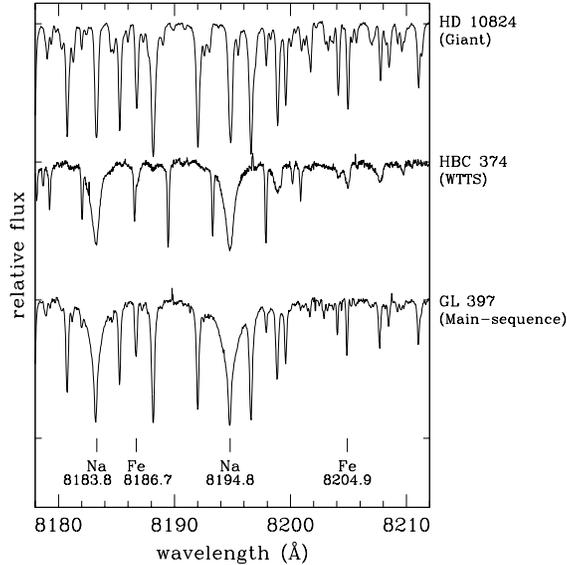}
   \end{center}
   \caption{Sample spectra of a giant, a WTTS, and a main-sequence star. The Doppler shift of each object is corrected. The absorption lines used to derive the EWRs are marked. Other sharp lines are mostly telluric lines. The Na (8183.3$\AA$) and Fe (8186.7$\AA$) in the WTTS spectrum are blended with the telluric lines. }\label{somelabel}
\end{figure}

The SPLOT task was used to measure the equivalent width (EW) of each line, and the line was fit with the Voigt function. Since telluric lines appear in this wavelength range, we de-blended these lines from the photospheric lines. The EWs of some Na and Fe lines were difficult to measure because of heavy telluric line blending. The error in EW was estimated from uncertainty of the continuum level. The measured EWs of each object are listed in table 2. 

\begin{longtable}{lrrrr}
   \caption{The list of the measured EW for each object.}
\hline 
\multicolumn{1}{c}{Name} & \multicolumn{4}{c}{EW ($\AA$)} \\
\cline{2-5} 
 & Na (8183.3$\AA$) & Fe (8186.7$\AA$) & Na (8194.8$\AA$) & Fe (8204.9$\AA$) \\
\hline
\endhead
\hline
\endfoot
\hline
\multicolumn{4}{@{}l@{}}{\hbox to 0pt{\parbox{180mm}{\footnotesize
       The lines heavily blended with telluric lines were unmeasured (expressed "---").
       }\hss}}
\endlastfoot
Giants &  &  &  &  \\
\hline
   $\beta$ Gem	& 0.280$^{+0.009}_{-0.011}$	& 0.100$\pm$0.006	& 0.374$^{+0.008}_{-0.009}$	& 0.084$\pm$0.001	\\
   HD 19656	& 0.264$^{+0.014}_{-0.010}$	& \multicolumn{1}{c}{---}	& 0.337$^{+0.016}_{-0.007}$	& 0.104$\pm$0.003	\\
   HD 76291	& 0.251$\pm$0.006	& 0.074$^{+0.007}_{-0.003}$	& 0.342$\pm$0.008	& 0.091$^{+0.003}_{-0.002}$	\\
   HD 100204	& 0.247$^{+0.007}_{-0.005}$	& 0.114$^{+0.007}_{-0.008}$	& 0.359$^{+0.007}_{-0.009}$	& 0.104$\pm$0.002	\\
   HD 145148	& 0.313$^{+0.006}_{-0.005}$	& 0.072$\pm$0.001	& 0.431$\pm$0.006	& 0.069$\pm$0.002	\\
   HD 45512	& 0.304$^{+0.014}_{-0.011}$	& 0.113$^{+0.006}_{-0.005}$	& 0.434$\pm$0.008	& 0.111$\pm$0.003	\\
   HD 101978	& 0.279$^{+0.006}_{-0.008}$	& 0.165$^{+0.006}_{-0.005}$	& 0.347$^{+0.012}_{-0.013}$	& 0.149$\pm$0.008	\\
   HD 106102	& 0.341$^{+0.011}_{-0.006}$	& 0.130$^{+0.001}_{-0.003}$	& 0.496$^{+0.017}_{-0.019}$	& 0.109$^{+0.001}_{-0.002}$	\\
   HD 108299	& 0.274$^{+0.007}_{-0.006}$	& 0.138$\pm$0.001	& 0.329$^{+0.007}_{-0.006}$	& 0.130$^{+0.004}_{-0.005}$	\\
   HD 110501	& 0.315$\pm$0.006	& 0.105$^{+0.005}_{-0.009}$	& 0.460$^{+0.013}_{-0.012}$	& 0.086$\pm$0.002	\\
   HD 121146	& \multicolumn{1}{c}{---}	& 0.126$^{+0.006}_{-0.005}$	& 0.393$^{+0.002}_{-0.001}$	& 0.103$^{+0.001}_{-0.002}$	\\
   HD 147142	& 0.293$^{+0.006}_{-0.005}$	& 0.109$\pm$0.005	& 0.393$\pm$0.008	& 0.122$^{+0.000}_{-0.001}$	\\
   HD 49738	& 0.309$\pm$0.011	& 0.129$^{+0.007}_{-0.005}$	& 0.375$^{+0.004}_{-0.005}$	& 0.150$\pm$0.001	\\
   HD 65759	& 0.293$^{+0.009}_{-0.008}$	& 0.114$\pm$0.001	& 0.360$^{+0.010}_{-0.007}$	& 0.136$^{+0.000}_{-0.001}$	\\
   HD 86369	& \multicolumn{1}{c}{---}	& 0.145$^{+0.006}_{-0.005}$	& 0.381$^{+0.007}_{-0.008}$	& \multicolumn{1}{c}{---}	\\
   HD 88231	& 0.291$\pm$0.013	& 0.126$\pm$0.004	& 0.368$^{+0.014}_{-0.015}$	& 0.131$\pm$0.001	\\
   HD 102328	& 0.374$^{+0.005}_{-0.012}$	& 0.150$\pm$0.003	& 0.542$\pm$0.012	& 0.134$^{+0.002}_{-0.003}$	\\
   HD 109012	& 0.289$\pm$0.006	& \multicolumn{1}{c}{---}	& 0.383$^{+0.010}_{-0.012}$	& 0.108$\pm$0.003	\\
   HD 113637	& 0.286$^{+0.003}_{-0.002}$	& 0.115$^{+0.008}_{-0.004}$	& 0.358$^{+0.004}_{-0.007}$	& 0.142$^{+0.004}_{-0.006}$	\\
   HD 115478	& 0.294$^{+0.002}_{-0.001}$	& 0.124$^{+0.003}_{-0.002}$	& 0.432$^{+0.004}_{-0.002}$	& 0.125$^{+0.004}_{-0.005}$	\\
   HD 118839	& \multicolumn{1}{c}{---}	& \multicolumn{1}{c}{---}	& 0.377$^{+0.009}_{-0.010}$	& 0.112$\pm$0.002	\\
   HD 129245	& \multicolumn{1}{c}{---}	& \multicolumn{1}{c}{---}	& 0.398$\pm$0.006	& 0.127$\pm$0.004	\\
   HD 132304	& \multicolumn{1}{c}{---}	& 0.132$\pm$0.002	& 0.300$^{+0.003}_{-0.004}$	& 0.120$\pm$0.003	\\
   HD 137688	& 0.248$^{+0.002}_{-0.003}$	& 0.113$^{+0.007}_{-0.003}$	& 0.350$^{+0.013}_{-0.010}$	& 0.124$^{+0.002}_{-0.004}$	\\
   HD 146388	& 0.277$^{+0.004}_{-0.005}$	& 0.110$\pm$0.007	& 0.394$\pm$0.001	& 0.106$\pm$0.001	\\
   HD 156093	& 0.263$^{+0.012}_{-0.016}$	& 0.142$^{+0.003}_{-0.002}$	& 0.343$\pm$0.001	& 0.131$\pm$0.001	\\
   HD 163547	& 0.264$^{+0.006}_{-0.008}$	& 0.122$^{+0.007}_{-0.005}$	& \multicolumn{1}{c}{---}	& 0.124$^{+0.004}_{-0.002}$	\\
   HD 10824	& 0.320$^{+0.009}_{-0.008}$	& 0.174$^{+0.001}_{-0.002}$	& 0.398$\pm$0.011	& 0.167$^{+0.006}_{-0.005}$	\\
   HD 20644	& 0.347$^{+0.016}_{-0.014}$	& 0.164$\pm$0.004	& 0.407$^{+0.016}_{-0.007}$	& 0.166$\pm$0.005	\\
   HD 23413	& 0.322$^{+0.010}_{-0.006}$	& \multicolumn{1}{c}{---}	& \multicolumn{1}{c}{---}	& 0.139$^{+0.009}_{-0.006}$	\\
\hline
Main-sequence stars & & & & \\
\hline
   GL 105	& 0.478$^{+0.013}_{-0.014}$	& 0.070$^{+0.005}_{-0.004}$	& 0.596$\pm$0.004	& 0.046$\pm$0.001	\\
   GL 183	& 0.629$^{+0.008}_{-0.010}$	& \multicolumn{1}{c}{---}	& \multicolumn{1}{c}{---}	& 0.065$^{+0.007}_{-0.006}$	\\
   GL 141	& 0.826$^{+0.031}_{-0.015}$	& 0.111$^{+0.007}_{-0.006}$	& 1.171$^{+0.020}_{-0.023}$	& 0.073$^{+0.001}_{-0.002}$	\\
   GL 380	& \multicolumn{1}{c}{---}	& 0.097$^{+0.003}_{-0.004}$	& 1.345$^{+0.048}_{-0.052}$	& 0.066$^{+0.004}_{-0.002}$	\\
   GL 397	& 0.940$^{+0.017}_{-0.015}$	& 0.117$\pm$0.011	& 1.297$^{+0.035}_{-0.028}$	& 0.069$\pm$0.002	\\
   GJ 3678	& 0.506$^{+0.009}_{-0.007}$	& 0.058$^{+0.008}_{-0.006}$	& 0.671$^{+0.012}_{-0.006}$	& 0.036$^{+0.002}_{-0.001}$	\\
\hline
WTTS & & & & \\
\hline
   RX J0452.5+1730 & \multicolumn{1}{c}{---}	& \multicolumn{1}{c}{---}	& 0.772$^{+0.026}_{-0.033}$	& 0.082$^{+0.015}_{-0.013}$	\\
   RX J0459.7+1430 & \multicolumn{1}{c}{---}	& \multicolumn{1}{c}{---}	& 0.720$^{+0.033}_{-0.065}$	& 0.086$^{+0.024}_{-0.017}$	\\
   HBC 374	   & \multicolumn{1}{c}{---}	& \multicolumn{1}{c}{---}	& 0.963$^{+0.037}_{-0.026}$	& 0.085$\pm$0.007	\\ 
   V827 Tau	   & \multicolumn{1}{c}{---}	& \multicolumn{1}{c}{---}	& 1.223$^{+0.021}_{-0.053}$	& 0.109$^{+0.020}_{-0.017}$	\\ \hline
   \end{longtable}


\section{Results and Discussion}

\subsection{Relationship between EWR and Surface Gravity}

We derived a relationship between log $g$ and the EWR of Fe and Na absorption lines (Fe/Na) for 25 giants and the 4 main-sequence stars of which $T_{\mathrm{eff}}$ are around 4200-4300K. We calculated the EWR using four line pairs: Fe (8186.7$\AA$) / Na (8183.3$\AA$), Fe (8186.7$\AA$) / Na (8194.8$\AA$), Fe (8204.9$\AA$) / Na (8183.3$\AA$), and Fe (8204.9$\AA$) / Na (8194.8$\AA$). Results are shown in figure 3. Filled squares and triangles mark the EWRs of the giants and main-sequence stars, respectively. The solid line in each graph is the best-fit line for these objects. In every case, the line ratios decrease significantly with increasing log $g$. The averaged slope of the curves in log $g$ = 3.5-4.0 (a typical surface gravity for PMS stars) are -0.085, -0.067, -0.084, -0.064 in Fe (8186.7$\AA$) / Na (8183.3$\AA$), Fe (8186.7$\AA$) / Na (8194.8$\AA$), Fe (8204.9$\AA$) / Na (8183.3$\AA$), and Fe (8204.9$\AA$) / Na (8194.8$\AA$), respectively. These EWRs are considered to be an excellent diagnostics for deriving the surface gravity. 

Next, we calculated the EWRs of five giants and two main-sequence stars of which $T_{\mathrm{eff}}$ is 4600-4700K. The EWRs of these objects are shown as open squares and triangles in figure 3. The best-fit line in each EWR is indicated as a dashed line. The EWRs increase for higher-$T_{\mathrm{eff}}$ objects at constant gravity over the range of log $g$ = 3.5-4.0. A comparison of fitted curves indicates that the EWRs of high-$T_{\mathrm{eff}}$ objects are 0.007-0.032 larger than the low-$T_{\mathrm{eff}}$ EWRs in log $g$ = 3.5.

\begin{figure}
   \begin{center}
      \FigureFile(160mm,140.9mm){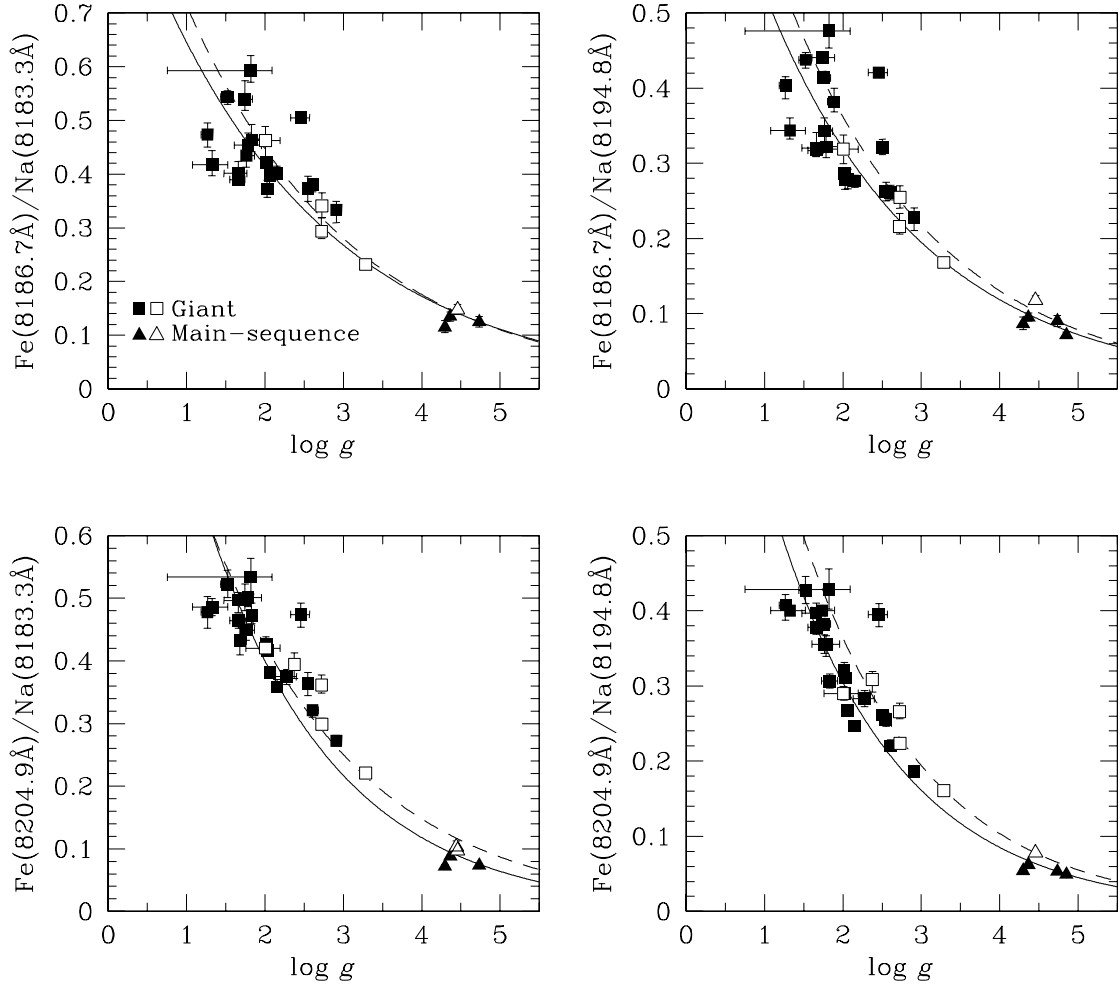}
   \end{center}
   \caption{EWR versus log $g$ ($g$: cm/$s^2$). The squares mark the EWRs for the giants. The triangles mark the EWRs for main-sequence stars. The filled plots denote the EWRs of objects with $T_{\mathrm{eff}}$ around 4200-4300K, and the open plots are for 4600-4700K objects. The solid lines are the fitted curve for the 4200-4300K objects. The lines for the 4600-4700K objects are shown as dashed. }\label{somelabel}
\end{figure}

\subsection{Model Atmospheric Spectra Comparisons}

We synthesized spectra using the SPTOOL program \citep{Takeda1995}, which calculates model spectra based on the ATLAS 9 model \citep{Kurucz1993}. The model spectra were calculated for $T_{\mathrm{eff}}$ of 4200K and 4750K, and a range in log $g$ of 1.0-5.0, and the EWRs were estimated (figure 4). All of the EWRs calculated from the synthesized spectra decreased with increasing log $g$. This is consistent with the EWRs for the observed spectra. 

In addition, although the EWRs of the observed spectra and the data from the SPTOOL are poorly matched especially in the EWRs from the 4750K model spectra and the Fe (8204.9$\AA$) ratios, the EWRs of the high-$T_{\mathrm{eff}}$ objects are larger than those of low-$T_{\mathrm{eff}}$ objects with a log $g$ range larger than 3. This occurs because the EW of the Fe lines increases, and the wing of the Na lines decreases in high-$T_{\mathrm{eff}}$ spectra. 

This discrepancy in observed spectra and model spectra could arise from improper values of the excitation potential and the damping constant of the Fe and Na lines. Since the both observed and synthetic EWRs match in the Fe (8186.7$\AA$) / Na (8183.3$\AA$) and Fe (8186.7$\AA$) / Na (8194.8$\AA$) EWRs, the excitation potential and the damping constant of Fe (8186.7$\AA$) line and both Na lines are seem to be estimated correctly. In the Fe (8204.9$\AA$) case, since the observed EWs were weaker than the modeled EWs and has no strong wings, the excitation potential may be underestimated. Further improvements on the line data will increase the utility of this log $g$ diagnostic.

\begin{figure}
   \begin{center}
      \FigureFile(160mm,142.2mm){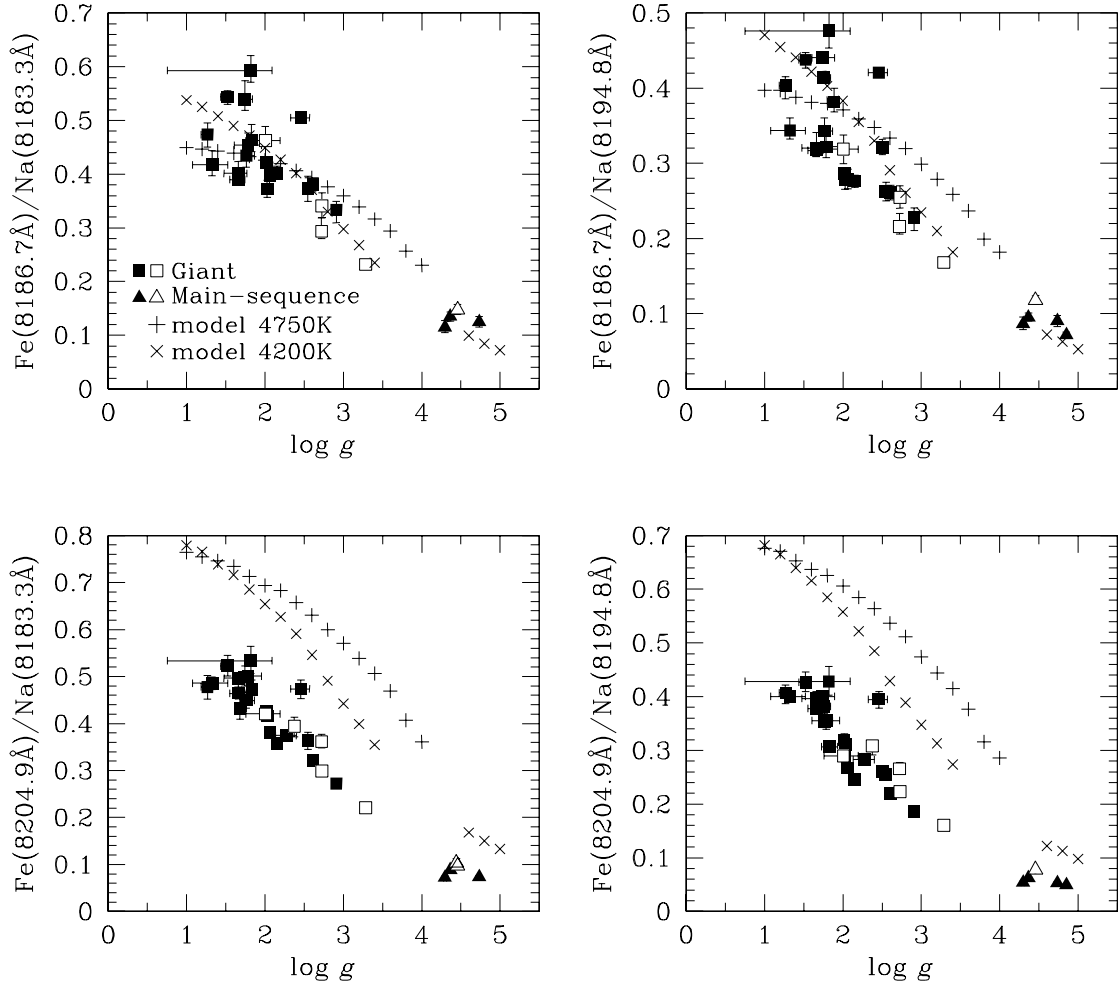}
   \end{center}
   \caption{EWR versus log $g$ ($g$: cm/$s^2$) with the EWR of the model spectra created by SPTOOL. The squares indicate the EWR of the giants, and the triangles represent the main-sequence stars. The filled plots denote the EWRs of the 4200-4300K objects, and the open plots are for 4600-4700K objects. The plus and cross signs are the EWR from model spectra of $T_{\mathrm{eff}}$ = 4750K and $T_{\mathrm{eff}}$ = 4250K, respectively.}\label{somelabel}
\end{figure}

\subsection{The EWR of WTTS}

We next compared the EWRs of WTTSs with the EWRs of giants and main-sequence stars. Since the Na (8183.3$\AA$) and Fe (8186.7$\AA$) lines of the WTTSs were blended with the telluric lines, only the Fe(8204.9$\AA$) / Na (8194.8$\AA$) was obtained. 

\begin{figure}
   \begin{center}
      \FigureFile(140mm,100mm){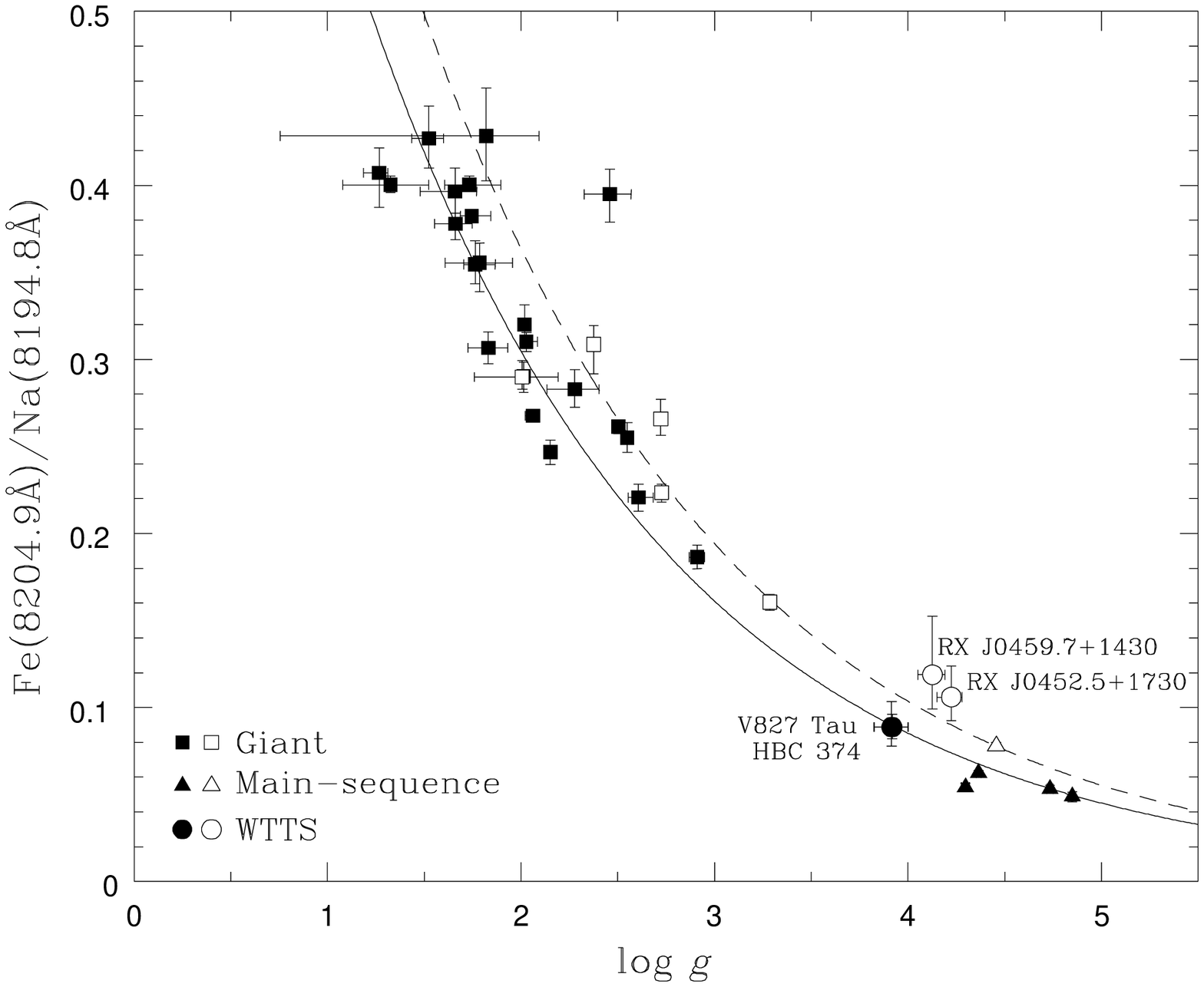}
   \end{center}
   \caption{EWR (Fe (8204.9$\AA$) / Na (8194.8$\AA$)) versus log $g$ ($g$: cm/$s^2$) for the EWR of WTTSs. The squares, triangles, and the fitted curves are from figure 3. The open and filled circles indicate the EWRs of WTTS of which $T_{\mathrm{eff}}$ = 4590K and $T_{\mathrm{eff}}$ = 4060K, respectively.}\label{somelabel}
\end{figure}

As seen in figure 5, the EWRs of two WTTSs (RX J0452.5+1730 and RX J0459.7+1430, marked with open circles) with $T_{\mathrm{eff}}$=4590K are larger than the EWR of the WTTS (HBC 374, marked with filled circle) of which $T_{\mathrm{eff}}$ is 4060K. This is consistent with results from the fitted curves and the model spectra mentioned in section 4.2. The EWRs of high-$T_{\mathrm{eff}}$ WTTSs agree closely with the curve from the 4600-4700K objects, and the EWRs of HBC 374 and V827 Tau fits the curve approximated by the 4200-4300K objects. This result indicates that the surface gravity of PMS stars is able to be determined from this relationship. 

By using this relationship, the log $g$ of these two WTTSs are estimated at 3.95, which correspond to the age of 2.8 Myr in the evolution model of \citet{Baraffe1998}. This result agrees with the age derived from the photometric observation using the extinction amount of \citet{Kenyon1995} (table 3). However, using the amount of extinction of \citet{Furlan2006} results the log $g$ of HBC 374 and V827 Tau to 3.263 and 3.627, respectively. The ages of these two WTTSs estimated by using this $A_V$ will be less than 1 Myr, which mismatch our result. 

\begin{table}
 \begin{center}
 \caption{The age of the WTTSs estimated from EWR and photometric method.}
  \begin{tabular}{lccccccc}
   \hline
   \multicolumn{1}{c}{Name}       & \multicolumn{7}{c}{Method} \\
   \hline
   			& \multicolumn{1}{c}{EWR}	&& \multicolumn{5}{c}{Photometric}	\\
   \cline{2-3} \cline{4-8}
   			& 				&& \multicolumn{2}{c}{KH95\footnotemark[$*$]}	&& \multicolumn{2}{c}{F06\footnotemark[$\dagger$]}	\\
   \cline{4-6} \cline{7-8}
   			& Age (Myr)	&& $A_V$	& Age (Myr)	&& $A_V$	& Age (Myr)	\\
   \hline
   HBC 374		& 2.8		&& 0.76	& 2.5		&& 2.40	& $<$1.0 	\\
   V827 Tau		& 2.8		&& 0.28	& 2.5		&& 1.00	& $<$1.0	\\
   \hline
    \multicolumn{8}{@{}l@{}}{\hbox to 0pt{\parbox{85mm}{\footnotesize
       \par\noindent
       \footnotemark[$*$] \citet{Kenyon1995}
       \par\noindent
       \footnotemark[$\dagger$] \citet{Furlan2006}
       }\hss}}
   \end{tabular}
 \end{center}
 \end{table}

\subsection{Age Determination Accuracy}

The age of a PMS star is able to be estimated by comparing the derived surface gravity with an evolution model (\cite{Baraffe1998}; \cite{Siess2000}). In our method, the accuracy of the calculated ages depends on the precision of the surface gravity value obtained for the PMS stars. This is dependent on the error in the EWR. We assumed the EWR error of the PMS star as $7.0\times10^{-3}$, which nearly corresponds to the error of HBC 374. The log $g$ error calculated from this uncertainty is 0.1-0.2 in the 3.5-4.0 range. Consequently, by comparing with an evolution model of a 0.8 $M_{\odot}$ star in \citet{Baraffe1998}, the age of the PMS star can be determined within a factor of 1.5, in average. This precision will be improved by deriving the general EWR error in the other three EWRs. To adapt this method to PMS stars of different temperatures, accurate estimations of the relation of EWR-log $g$ relationship for each $T_{\mathrm{eff}}$ range are required. In addition, an independent $T_{\mathrm{eff}}$ estimation is necessary in order to determine the log $g$ of the pre-main sequence star using this technique.


\section{Conclusion}

We carried out high-resolution optical spectroscopy to create a new age determination method for late-K type PMS stars. To obtain a surface gravity indicator from the spectrum, we derived the EWR using nearby absorption lines, which avoid problems from veiling contamination and distance uncertainty. 

The EW of the Na (8183.3$\AA$ and 8194.8$\AA$) and Fe (8186.7$\AA$ and 8204.9$\AA$) absorption lines in the $I$-band was used in this work. To derive the relationship between Fe/Na and surface gravity, we observed 25 giant stars, 4 main sequence stars, and WTTSs with the Okayama Astrophysical Observatory 1.88 m telescope and HIDES, and the Subaru Telescope and HDS. The $T_{\mathrm{eff}}$ of the objects were 4200-4300K and the surface gravity was known from the previous photometric studies. All four EWRs decreased with increasing surface gravity. Therefore, they are efficient diagnostics of surface gravity. Using the Fe (8204.9$\AA$) / Na (8194.8$\AA$) EWR, the log $g$ of PMS stars with 0.8 $M_{\odot}$ can be estimated with an uncertainty of 0.1-0.2. From comparisons with an evolution model, we concluded that their ages can be determined within a factor of 1.5.

Five giants, two main sequence stars, and two WTTSs with higher-$T_{\mathrm{eff}}$ (4600-4700K) were also observed to estimate the temperature dependence of the EWR. The EWRs of the high-$T_{\mathrm{eff}}$ objects were larger than those of the low $T_{\mathrm{eff}}$ in the typical PMS star surface gravity range. Therefore, to accurately determine the ages of PMS stars with masses larger or smaller than 0.8 $M_{\odot}$, new EWR-log $g$ relationship for each $T_{\mathrm{eff}}$ range are required. A precise estimation of EWR-log $g$ relationship allow us to determine the accurate age of PMS stars and to study the evolution process of star and circumstellar materials. 

~\\
~\\

We are grateful for the assistance given by the staff of the Okayama Astrophysical Observatory during the observations. We also thank the staff of the Subaru Telescope for the service observation.

\end{document}